\documentclass[12pt,twoside]{article}
\usepackage{amssymb}
\usepackage{amsthm}
\usepackage{graphicx}
\def\R{{\mathbb R}}
\def\N{{\mathbb N}}

\newtheorem{thm}{Theora}[section]

\newtheorem{theo}[thm]{Theorem}

\newtheorem{Def}[thm]{Definition}

\newtheorem{ex}[thm]{Example}

\def\prend{$~~\mbox{\hfil\vrule height6pt width5pt depth-1pt}$ }
\begin{document}
\pagestyle{myheadings} \markboth{ S. H. Djah, H. Gottschalk, H.
Ouerdiane }{Feynman graphs for general measures}
\thispagestyle{empty}

\title{Feynman graphs for non-Gaussian measures}
 \author{Sidi Hamidou Djah${}^1$, Hanno Gottschalk${}^2$ and Habib Ouerdiane${}^1$}
\maketitle {\small

 \noindent ${}^1$: D\'epartement des Math\'ematiques, Universit\'e de Tunis El Manar

 \noindent  ${}^2$: Institut f\"ur angewandte Mathematik, Rheinische Fridrich-Wilhelms-Universit\"at Bonn

\vspace{1cm}

\noindent{\bf Abstract.}  Partition and moment functions for a
general (not necessarily Gaussian) functional measure that is
perturbed by a Gibbs factor are calculated using generalized
Feynman graphs. From the graphical calculus, that is formulated using the
theory of species, a new notion of Wick
ordering arises, that coincides with orthogonal decompositions of
Wiener-It\^o ~type only if the measure is Gaussian. Proving a
generalized linked cluster theorem, we show that the logarithm of
the partition function can be expanded in terms of connected
Feynman graphs.



\vspace{.2cm}

\section{Introduction}
Perturbative expansions for Path integrals -- both oscillatory
\cite{Fe,AHK} and probabilistic \cite{Mi,It,Si,GJ} are one of the
main technical tools in contemporary elementary particle physics
and statistical physics.

Expansions in terms of Feynman graphs are done starting of with a
free measure of Gaussian type which is then perturbed with a local
and polynomial interaction, e.g. the $\phi^4$ interaction.

In this note we give some results on the perturbation theory of
non-Gaussian measures of Euclidean type. The focus is on the
combinatorial expansion in terms of a generalized class of graphs,
henceforth called generalized Feynman graphs.

Non Gaussian measures as a starting point of the theory are of
interest in statistical physics, where e.g. measures of poisson
type play a great r\^ole \cite{Ru}, but recently also have
attracted some attention in particle physics, cf. e.g. \cite{Kl}, and
-- properly generalized to Grassmann algebras -- in the theory of solid states \cite{Br}.
Physically motivated questions related to infra-red and
self energy problems of systems of classical particles in the continuum can be found in \cite{DGO}.
Also in stochastics, related graph expansions proved to be useful in the asymptotic expansion of densities of L\'evy laws \cite{GST} and
the perturbation theory of L\'evy driven SPDEs \cite{GS}.

In particular, a reformulation of the Wilson-Polchinski renormalization group for measures other than Gaussian
seems to be highly desirable, since only if different systems of statistical mechanics can be treated on the same mathematical footing,
universality can be properly understood, cf. \cite{GOS} for a first step in that direction. It is clear, that all these works heavily rely on
combinatorial structures as a point of departure.

In previous articles on the subject, cf. \cite{DGO,GOS}, the Feynman graph calculus was however
used as in theoretical physics, as an {\em ad hoc} formalism
 to order certain combinatorial sums. In this work we refine the definition
of the main cass of objects -- the generalized Feynman graphs -- using the proper
combinatorial tools given by the theory of species \cite{BLL}.

After a short section fixing the notations and briefly surveying
the perturbation expansion, we introduce these generalized Feynman
graphs in section 3. In section 4 we give a graphical notion of
Wick ordering for general functional measures and we compare it
with orthogonal expansions of Wiener-It\^o-Segal type
\cite{IK,Si,GJ}. In section 5 we prove the linked cluster theorem
for the generalized Feynman graphs, which paves the way for
applications in statistical physics.

\section{Perturbation theory}
Let us consider the nuclear triplet
\begin{equation}
\label{2.1eqa}{\cal S}(\R^d)\subset{L^2(\R^d)}\subset{{\cal
{S'}}(\R^d)}
\end{equation}where $d\in \N,\; {\cal S}(\R^d)$ the space of
functions defined on $\R^d,$ differentiable for any order, rapidly
decreasing and ${\cal {S'}}(\R^d)$ the space of tempered
distributions, i.e the  topological dual of ${\cal S}(\R^d).$ We
denote by ${\cal B}({\cal {S'}}(\R^d))$ the Borel $\sigma-$algebra
generated by the open sets of the weak topology on ${\cal
{S'}}(\R^d)$.\\ We consider functional measures $\nu_0$ on the
measurable space $({\cal {S'}}(\R^d), {\cal B}({\cal
{S'}}(\R^d)))$ such that all moments of $\nu_0$ exist. Let
$\chi_{\epsilon}\in{\cal S}(\R^d)$ be a family of positive,
reflection invariant function such that
$\int_{\R^d}\chi_{\epsilon}(x) dx=1,$
$\chi_{\epsilon}\rightarrow\delta_0$ in ${\cal {S'}}(\R^d)$ for
$\epsilon\downarrow 0$. The function $\chi_{\epsilon}$ can be seen
as the integral kernel of an operator :
\begin{equation}\label{2.2eqa}
\chi_{\epsilon}:\;{\cal{S'}}(\R^d)\rightarrow
{\cal{S'}}(\R^d)\;,\; \;  \phi\mapsto
\phi_{\epsilon}=\chi_{\epsilon}*\phi
\end{equation}where $(\chi_{\epsilon}*\phi)(x)=\langle \phi,\chi_{\epsilon,x}
\rangle,\; \; \chi_{\epsilon,x}(y)=\chi_{\epsilon}(y-x).$\\ The
convolution $\chi_{\epsilon}*\phi$ is continuous in the weak
${\cal{S'}}(\R^d)$ topology and in particular it is ${\cal
B}({\cal {S'}}(\R^d))-{\cal B}({\cal {S'}}(\R^d))$ measurable. It
takes values in the $C^{\infty}(\R^d)-$functions with tempered (at
most polynomial) increase at infinity,
$C^\infty_{temp}(\R^d)$.\\We consider the functional measure
$\nu_{0,\epsilon}$ which is the image measure of $\nu_0$ under
\\$\chi_{\epsilon}:{\cal {S'}}(\R^d)\rightarrow {\cal
{S'}}(\R^d).$ In particular, the support of $\nu_{0,\epsilon}$
lies in the image of ${\cal{S'}}(\R^d)$ under $\chi_{\epsilon},$
hence in the space of $C^{\infty}_{temp}(\R ^d)$.
 It is thus possible, to approximate (in law) any measure
$\nu_0$ by measures $\nu_{0,\epsilon}$ which have support not only
on tempered distributions, but on
$C^{\infty}_{temp}(\R^d)-$functions. This goes under the name
ultra-violet (uv) regularization procedure.\\ From now on we
assume that $\nu_0$ is suitably uv-regularized, i.e. has support
on $C^{\infty}_{temp}(\R^d).$ As a measurable space,
$C^\infty_{temp}(\R^d)$ is equipped with the trace $\sigma$
algebra ${\cal B}=C^{\infty}_{temp}(\R^d)\bigcap{\cal B}({\cal
{S'}}(\R^d))$.\\We can note that for $x\in \R^d$, the mapping
$C^{\infty}_{temp}(\R^d)\ni \phi\mapsto \phi(x)\in\R$ is
measurable as the pointwise limit of $\langle \phi,
\chi_{\epsilon, x}\rangle\rightarrow \phi(x)\; \; \;
\epsilon\downarrow 0$. Having specified the conditions on $\nu_0,$
we next define interactions of local type :\\Let $v: \R\rightarrow
\R$ be a function which is continuous and bounded from below. Let
$\Lambda$ be a compact subset of $\R^d.$ Then we set
\begin{equation}\label{2.3eqa}V_{\Lambda}(\phi)=\int_{\Lambda}
v(\phi(x))dx,\; \; \phi \in C^{\infty}_{temp}(\R^d) \end{equation}
 We are particularly
interested in the case where $v$ is a monomial and we chose
$v(\phi)=\phi^4$ for the sake of concreteness following the
tradition in perturbation theory. Furthermore, it is easy to show
that $V_{\Lambda} : C^{\infty}_{temp}(\R^d)\rightarrow \R$ is
${\cal B}-$measurable and bounded from below. Hence, $V_\Lambda,
e^{-\lambda V_\Lambda}\in L^p(\nu_0)$ for $\lambda>0,p\geq1$ by
the assumption on existence of moments and semiboundedness.
\\ Let $\nu_{\Lambda}^{\lambda}$ be the non-normalized perturbed
measure defined by
:\begin{equation}\label{2.4eqa}d\nu_{\Lambda}^{\lambda}(\phi)=
e^{-\lambda V_{\Lambda}(\phi)}d\nu_0(\phi)\end{equation} where
$0<\lambda \ll 1.$ $\lambda$ is called the coupling constant.\\ We
want to calculate the moments of $\nu_{\Lambda}^{\lambda} :$
\begin{equation}\label{2.5eqa} S_{n,\Lambda}^{\lambda}(x_1,...,x_n)=
\int_{C^{\infty}_{temp}}\phi(x_1)...\phi(x_n) e^{-\lambda
V_{\Lambda}(\phi)}d\nu_0(\phi)\end{equation} \\  Next we come to
the perturbation series of the expression (\ref{2.5eqa}). One can
expand the exponential in (\ref{2.5eqa}) in powers of the coupling
constant :\begin{equation}\label{2.6eqa}
S_{n,\Lambda}^{\lambda}(x_1,...,x_n)=
\int_{C^{\infty}_{temp}}\phi(x_1)...\phi(x_n)\sum_{m=0}^{\infty}
{(-\lambda)^m\over{m!}}\left[\int_{\Lambda}\phi^4(x)dx\right]^md\nu_0(\phi)\end{equation}
Next one interchanges the infinite sum and the integral w.r.t
$\nu_0.$ This is in general only a formal operation
:\begin{equation}\label{2.7eqa}
S_{n,\Lambda}^{\lambda}(x_1,...,x_n)=\sum_{m=0}^{\infty}
{(-\lambda)^m\over{m!}}\int_{C^{\infty}_{temp}}\phi(x_1)...\phi(x_n)\left[\int_{\Lambda}
\phi^4(x)dx\right]^md\nu_0(\phi)\end{equation}Applying Fubini's
lemma to the right hand side of (\ref{2.7eqa}) one obtains
\begin{equation}\label{2.8eqa}
S_{n,\Lambda}^{\lambda}(x_1,...,x_n)=\sum_{m=0}^{\infty}
{(-\lambda)^m\over{m!}}\int_{{\Lambda}^m}\int_{C^{\infty}_{temp}}
\phi(x_1)...\phi(x_n)\phi^4(y_1)...
\phi^4(y_m)d\nu_0(\phi)dy_1...dy_m
\end{equation}
The problem with (\ref{2.8eqa}) is that for a number of
interesting examples the series diverges, while (\ref{2.6eqa}) is
well-defined. From the semiboundedness of $V_\Lambda$ one obtains that
$S^\lambda_{n,\Lambda}$ is $C^\infty$ in $\lambda$ on $[0,+\infty)$ however not
necessarily analytic at 0, \cite[Lemma 2.2]{DGO}. Therefore, a finite
partial sum up to the order $N\in \N$ in the coupling constant $\lambda >
0$ of the divergent series (\ref{2.8eqa}) for $\lambda$ small
gives an excellent approximation of (\ref{2.6eqa}), i.e. for all
$N\in\N$ fixed :
\begin{equation}
\left|S_{n,\Lambda}^{\lambda}(x_1,...,x_n)-\sum_{m=0}^{N}
{(-\lambda)^m\over{m!}}\int_{{\Lambda}^m}\int_{C^{\infty}_{temp}}
\!\!\!\!\!\phi(x_1)...\phi(x_n)\phi^4(y_1)...
\phi^4(y_m)d\nu_0(\phi)dy_1...dy_m\right|<c_N\lambda^{N+1}
\end{equation}
 for
some $c_N>0$ by Taylor's lemma. Note that in particular the case
$n=0$ gives us an expression for the sum over states
\begin{equation}
S_{0,\Lambda}^\lambda=\Xi_\Lambda(\lambda)=
\int_{C^{\infty}_{temp}} e^{-\lambda
V_{\Lambda}(\phi)}d\nu_0(\phi)
\end{equation}
 and the moment functions for the
normalized measure $\Xi_\Lambda(\lambda)^{-1}\nu_\Lambda^\lambda$
can be obtained from the expansion on $S_{n,\Lambda}^\lambda$ and
$\Xi_\Lambda(\lambda)$ by the procedure of inversion and
multiplication of inverse power series. In section 5 we will also
consider the case of the free energy  $\ln \Xi_\Lambda(\lambda)$
which is the main object of interest in thermodynamics.\\ To
obtain the formal perturbation series, it is enough to evaluate
the integrals in the formal power series (\ref{2.8eqa}) :
$$
\int_{{ \Lambda}^m}\int_{C^{\infty}_{temp}}
\phi(x_1)...\phi(x_n)\phi^4(y_1)...
\phi^4(y_m)d\nu_0(\phi)dy_1...dy_m
$$
\begin{equation}\label{2.9eqa}=\int_{{
\Lambda}^m} \langle\phi(x_1)...\phi(x_n)\phi^4(y_1)...
\phi^4(y_m)\rangle_{\nu_0}dy_1...dy_m
\end{equation}
\\In the following section we give a graphical method for the
evaluation of such integrals. This method is standard for the
special case when $\nu_0$ is Gaussian and then gives the
well-known Feymann diagrams and rules. The generalization to
non-Gaussian measures is our objective.
\section
{Generalized Feynman graphs} The content of this and the following
sections is essentially combinatorial. Therefore give a short
digression on the theory of species \cite{BLL} which provides the
right language to deal with the partially labeled objects that are
needed in the following.

\begin{Def}
\label{3.0def} A species of structures is a rule $F$ such that for
any finite set $U$ another finite set $F[U]$ is generated and
furthermore for any bijection $\sigma: U\to V$ of finite sets a
mapping $F[\sigma]:F[U]\to F[V]$ is generated fulfilling the
following functorial properties
\begin{itemize}
\item[(i)] $F[I_U]=I_{F[U]}$ where $I$ stands for the identity mapping;
\item[(ii)] $F[\sigma]\circ F[\tau]=F[\sigma\circ\tau]$, for $U\stackrel{\tau}{\to}V\stackrel{\sigma}{\to}W$ bijections.
\end{itemize}
\end{Def}
Note that properties (i) and (ii) together imply that $F[\sigma]$ is a bijection, if $\sigma$ is.

Let us give some useful examples: For any set $U$ we may set
$I[U]=U$ and $I[\sigma]=\sigma$ and we trivially get a species.
Another species is the species of partitions ${\cal
P}[U]=\{I=\{I_1,\ldots,I_k\}:k\in{\N}^*, I_j\subseteq U, I_l\cap
I_j=\emptyset$ for $j\not=l$, $j,l=1,\ldots,k\}$ with ${\cal
P}[\sigma](I)=\{\sigma(I_1),\ldots,\sigma(I_k)\}$ if
$I=\{I_1,\ldots,I_k\}$.

For later use we note that the notion of species over one finite
set $U$ immediately generalizes zu ther notion of species over an
ordered pair of sets $(U,V)$ with isomorphisms
$\sigma=(\sigma_1,\sigma_2):(U,V)\to (W,Y)$ such  that
$\sigma_1:U\to W$ and $\sigma_2:V\to Y$ are bijections of finite
sets. We also introduce the notation $[n]$ for the set
$\{1,\ldots,n\}$ and $nU$ for $U\times[n]$.

We now come back to the point that in the
formula (\ref{2.8eqa}) in Section 2, the calculation of the
perturbation series can be reduced to the calculation of the
moments (\ref{2.9eqa}) of the unperturbed measure $\nu_0$. But
often it is easier to calculate truncated moments (also called truncated
Schwinger functions) which are defined as follows:
\begin{Def}\label{3.1Def}
Let $\nu_0$ be a probability measure defined on some measurable
space and let $X_1,X_2,\ldots$ be (not necessarily distinct)
random variables on that probability space that are $L^p$
integrable for $p\geq 1$. Then $\{\langle X_{j_1}\cdots
X_{j_n}\rangle_{\nu_0}\}$, $n\in{\N}^*$, $j_1\ldots,j_n\in\N$ is
the associated collection of moments. The truncated sequence of
monemts, $\langle X_{j_1},\ldots,X_{j_n}\rangle^T_{\nu_0}$ is then
defined recursively via
\begin{equation}\label{3.1eqa}{\langle
X_{j_1}\cdots X_{j_n}\rangle}_{\nu_0}=\sum_{\begin{array}{ll}I \in
{\cal P}(\{1,...,n\})\\ I=\{I_1,...,I_k\}\end{array}}
\prod_{l=1}^k{\langle I_l\rangle}^T_{\nu_0}
\end{equation} here $\langle
I_l\rangle_{\nu_0}^T=\langle
X_{j_{i_1}},\ldots,X_{j_{i_q}}\rangle^{T}_{\nu_0}$ for
$I_l=\{i_1,\ldots,i_q\}$.

In particular let $\nu_0$ be a measure on some space of functions and $X_j=\phi(u_j)$ for $u_1,u_2\ldots\in
\R^d$. The truncated moments $\langle
\phi(u_1),\ldots,\phi(u_n)\rangle^T_{\nu_0}$ of $\nu_0$ are
also called the truncated Schwinger functions of $\nu_0$. \end{Def} \noindent \\ The
following theorem is well-known:
\begin{theo}\label{3.1theo}(Linked Cluster theorem)\\
Let $C(t_1,u_1,...,t_n,u_n)\; , n\in \N \; , u_j
\in{\R^d},\;t_j\in \R$ be the Fourier transform of $\nu_0,$ i.e :
\begin{equation}\label{3.2eqa} C(t_1,u_1,...,t_n,u_n)=\int
e^{i\sum_{j=1}^{n}t_j \phi(u_j)}
d\nu_0(\phi)=\hat{{\nu}_0}(\sum_{j=1}^{n}t_j\delta_{u_j}).\end{equation}
Then the moments are generated by $C$
:\begin{equation}\label{3.3eqa}{\langle
\phi(u_1)...\phi(u_n)\rangle}_{\nu_0}=(-i)^n\frac{d^n}{dt_1...dt_n}\displaystyle
{C(t_1,u_1,...,t_n,u_n)|}_{t_1=...t_n=0}.\end{equation} Likewise,
let $C^T(t_1,u_1,...,t_n,u_n)=\ln C(t_1,u_1,...,t_n,u_n)$ which is
well-defined for $t_1,...,t_n$ sufficiently small. Then $C^T$
generates the truncated moments
:\begin{equation}\label{3.4eqa}{\langle
\phi(u_1),...,\phi(u_n)\rangle}^T_{\nu_0}=(-i)^n\frac{d^n}{dt_1...dt_n}C^T(t_1,u_1,...,t_n,u_n)
|_{t_1=...t_n=0}\end{equation}
\end{theo}
\begin{ex}{\rm (The Gaussian case)
If $\nu_0$ is Gaussian measure with covariance $G(u,v)$ and mean
$0$, then :$$C(t_1,u_1,...,t_n,u_n)=e^{-\frac{1}{2}
\displaystyle\sum_{j,\;l=1}^n t_j t_lG(u_j,u_l)}$$ and
consequently
\begin{equation}\label{3.5eqa}{\langle
\phi(u_1)\phi(u_2)\rangle}_{\nu_0}= G(u_1,u_2),
\end{equation}
\begin{equation}\label{3.6eqa}C^T(t_1,u_1,...,t_n,u_n)=-\frac{1}{2}\sum_{j,\;l=1}^n
t_j t_l G(u_j,u_l) \end{equation} Inserting this into
(\ref{3.4eqa}) yields :
\begin{equation}\label{3.7eqa}{\langle\phi(u_1),...,\phi(u_n)\rangle}_{\nu_0}^T=\left
\{\begin{array}{l} {0,\;n\neq 2}\\\\{ G(u_1,u_2),\; n=2},
\end{array}\right.\end{equation}
Using the linked Cluster theorem for this special case, one
obtains from ( \ref{3.1eqa}):
\begin{equation}\label{3.8eqa}{\langle
\phi(u_1)...\phi(u_n)\rangle}_{\nu_0}= \sum_{I\in {\cal
P}_e(\{1,...,n\})}\prod_{\{j,\;l\}\in I}G(u_j,u_l),\end{equation}
where ${\cal P}_e(\{1,...,n\})$ is the set of all partitions of
$\{1,...,n\}$ given by $I=\{I_1,...,I_{\frac{n}{2}}\}$ such that $|I_l|=2$ if $n$ is
even. Note that ${\cal P}_e(\{1,...,n\})=\emptyset$ if $n$ is odd.
Although known before \cite{Is}, the above formula in the physics context is known as Wick's theorem.}
\end{ex}
Now one can use the
combinatorics of truncation to obtain an expression of
(\ref{2.9eqa}) in terms of the usually simpler truncated moments.
In fact by Definition~\ref{3.1Def}, (\ref{2.9eqa}) is given by
:\begin{equation}\label {3.9eqa}\int_{{
\Lambda}^m}
\langle\phi(x_1)...\phi(x_n)\phi^4(y_1)...
\phi^4(y_m)\rangle_{\nu_0}dy_1...dy_m=\sum_{I\in{\cal
P}([n]+4[m])}\int_{{ \Lambda}^m}\prod_{I_l\in I}\langle I_l
\rangle^T_{\nu_0}dy_1...dy_m,\end{equation}
 where $\langle I_l\rangle^T_{\nu_0}=\langle\phi(z_{j_1}),\ldots, \phi(z_{j_q})\rangle^T_{\nu_0}$ with
$z_{j_l}=x_{j_l}$ for $j_l\in[n]$ and $z_{j_l}=y_s$ for $j_l\in 4[m]$, $j_l=(s,q)$, $s=1,\ldots,m$, $q=1,\ldots,4$, $I_l=\{j_1,\ldots j_q\}$.
In fact, given finite sets $U,V$ we could now define
$F[U,V]={\cal P}(U\dot\cup 4V)$ and {\em define} this as the species of generalized Feynman graphs
  with exterior (full) vertices labeled by $U$ and interaction (inner full) vertices labeled by $V$. For $|U|=n$ and $|V|=m$, one can then define an evaluation mapping $\hat {\cal V}_\Lambda(I)(x_{1},\ldots,x_{n})$ for $I\in F(U,V)$ by the integral on the right hand side of (\ref{3.9eqa}). In fact, as our calculations in the subsequent sections will show, this would be the best definition for proving theorems. It is however a bad definition for finding theoems, since this definition is not very
intuitive.

To visualize the combinatorics one thus uses a
representation through graphs, which is more intuitive. Now there are two problems. Firstly, what does "a representation through graphs" mean? Secondly, the standard notion of graphs has labeled vertices and non labeled "legs" (endpoints where the edges meet the vertex). This is adequate to formulate a graphic representation of the Meyer series for the low activity expansion in the staistical mechanics of gases and liquids \cite{Ru}.
It is however inadequate for the Feynman, high temperature expansion.
It will turn out that there we need
 vertices which are labeled and have labeled "legs" as well as vertices
 which are unlabeled and have unlabeled "legs" where the term "leg" stands for
 the points where the edge meets the vertex.

We start with the second problem. Let $U$ be a finite set, then
${\cal G}[U]$ is the species of simple graphs over the set of
vertices $U$, i.e. ${\cal G}[U]=\{ G\subseteq \{\{a,b\}\subseteq
U:  a\not=b\}\}$. The elements $e\in G$ are called edges of $G$.
The transport along a bijection $U\stackrel{\sigma}{\to}V$ is
defined by ${\cal G}[\sigma](G)=\{\sigma(e):e\in G\}$. For a
vertex $u\in U$ and a graph $G\in {\cal G}[U]$ the degree  is
defined by $f(u,G)=|\{e\in G:u\in e\}|$.

For $K\subseteq U$, let $\Sigma(K)$ be the group of permutations
of $K$ that acts as the identity on $U\setminus K$. Let ${\cal
G}'\subseteq {\cal G}[U]$ be a set of simple graphs that is
invariant under all ${\cal G}[\sigma]$ with $\sigma\in \Sigma(K)$.
Then let ${\cal G}'/\Sigma(K)$ be the set of equivalence classes
of graphs in ${\cal G}'$ that are being mapped onto each other via
the action of ${\cal G}[\Sigma(K)]$. The idea is to make the
vertices in $K$ unlabeled while the vertices in $U\setminus K$
remain labeled.

\begin{Def}\label{3.2Def}
 A generalized Feynman graph $G$ of $\phi^4$-theory with a set $U$ outer full vertices, a set $V$ of inner full vertices and  a set $K$ of inner empty vertices is a
graph $G\in{\cal G}[U\dot\cup 4V\dot\cup K]/\Sigma[K]$  such that the following conditions hold:
\begin{itemize}
\item[(i)] For $v\in U\dot\cup 4V$, $f(v,\hat G)=1$ and, for $v\in
K$, $f(v,\hat G)>0$ where $\hat G$ is a representative for the
equivalence class $G$; \item[(ii)] If $\hat G$ represents the
equivalence class $G$, then $\forall e\in\hat G$, $e=\{a,b\}$ with
$a\in U\dot\cup 4I$, $b\in K$.
\end{itemize}
Denote the set of such graphs by ${\cal F}(U,V,K)$. The set of all Feynman graphs with outer full vertices $U$ and inner full vertices $V$ is defined as ${\cal F}(U,V)=\dot\cup_{k\geq0}{\cal F}(U,V,[k])$.
\end{Def}
\noindent Next we work out a graphical representation. We use the conventions:
\begin{equation}\label{3.10eqa}Full\; inner\; vertex: {\bullet} \; \; \;  Empty\; inner\;
vertex: \circ \; \; \; Full \; outer\; vertex: \times
\end{equation}
Note that in a $\phi^4$ theory, each inner full vertex is of
fertility four. A vertex $v\in V$ is therefore identified with the
collection of its legs, $(v,1),\ldots,(v,4)\in 4V$: 
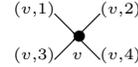
\hspace{.5cm}
\begin{picture}(1.5,.5)\put(.3,-.3){\line(1,1){.6}}
\put(.3,.3){\line(1,-1){.6}} \put(.53,-.1){$ \bullet$}
\put(.53,-.3){$\scriptscriptstyle v$}
\put(-.25,.3){$\scriptscriptstyle (v,1)$}
\put(.9,.3){$\scriptscriptstyle (v,2)$}
\put(-.25,-.3){$\scriptscriptstyle (v,3)$}
\put(.9,-.3){$\scriptscriptstyle (v,4)$}
\end{picture}
By condition (i),\\\vspace{-.2cm}

\noindent each leg is connected with exactly one edge. By condition (ii) full vertices are connected with empty vertices and vice versa, cf. the following examples:

\begin{ex}\label{1.2ex}
\end{ex}
\begin{picture}(8,3)
\thicklines \put(1,2){$\times$} \put(6.7,2){$\times$}
\put(3.05,2){$\bullet$} \put(4.85,2){$\bullet$}
\put(2.03,2){$\circ$} \put(5.83,2){$\circ$}
\put(3.94,1.5){$\circ$} \put(3.94,2.4){$\circ$}
\put(1.15,2.1){\line(1,0){.9}}
\put(6,2.1){\line(1,0){.85}}\put(1,2.9){$Generalized\;
Feynman\; Graph$} \put(6.7,.1){${\bf Figure\;1}$}
\bezier{100}(2.2,2.15)(2.6,2.45)(3.1,2.15)
\bezier{100}(2.2,2.05)(2.6,1.7)(3.1,2.05)
\bezier{100}(3.15,2.05)(3.6,1.65)(3.95,1.6)
\bezier{100}(3.15,2.15)(3.6,2.45)(3.95,2.5)
\bezier{100}(4.13,2.5)(4.5,2.45)(4.9,2.15)
\bezier{100}(4.13,1.6)(4.55,1.65)(4.9,2.05)
\bezier{100}(5,2.05)(5.4,1.65)(5.85,2.05)
\bezier{100}(5,2.15)(5.4,2.45)(5.85,2.15) \thicklines
\put(8,2){$\times$} \put(13.7,2){$\times$}
\put(10.05,2){$\bullet$} \put(11.85,2){$\bullet$}
\put(10.07,2.03){\line(0,-1){.8}} \put(12.83,2){$\circ$}
\put(10,1.03){$\circ$} \put(10.94,1.5){$\circ$}
\put(10.94,2.4){$\circ$} \put(8.15,2.1){\line(1,0){1.9}}
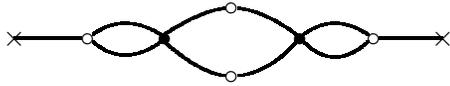
\put(13,2.1){\line(1,0){.85}}\put(8,2.9){No $Generalized\;
Feynman\; Graph$} \bezier{100}(10.15,2.05)(10.6,1.65)(10.95,1.6)
\bezier{100}(10.15,2.15)(10.6,2.45)(10.95,2.5)
\bezier{100}(11.13,2.5)(11.5,2.45)(11.9,2.15)
\bezier{100}(11.13,1.6)(11.55,1.65)(11.9,2.05)
\bezier{100}(12,2.05)(12.4,1.65)(12.85,2.05)
\bezier{100}(12,2.15)(12.4,2.45)(12.85,2.15)
\end{picture}

\noindent  The figure on right in Example~\ref{1.2ex} is no
Generalized Feynman graph, because there is an edge which connects
two full vertices. As it is customary, only the topological graph
(i.e. without labelings) is displayed.

We now come back to the first of the above mentioned problems. In what sense $F(U,V)$ and ${\cal F}(U,V)$ can be identified? The following definition helps:

\begin{Def}
\label{3.xdef}
 Two species $H$ and $R$ are equivalent, if for each finite set $U$ there is a bijection
 $\alpha[U]:H[U]\to R[U]$ such that for any finite set $V$ and any bijection $\sigma :U\to V,$
 the following diagram commutes
\begin{equation}
\label{3.x1eqa}
\begin{array}{ccc}
H[U]&\stackrel{\scriptscriptstyle\alpha[U]}{\longrightarrow}&R[U]\\
{\scriptscriptstyle H[\sigma]}\downarrow&&\downarrow{\scriptscriptstyle R[\sigma]}\\
H[V]&\stackrel{\scriptscriptstyle\alpha[V]}{\longrightarrow}&R[V]
\end{array}
\end{equation}
\end{Def}
Again, the generalization to species that depend on two finite
sets is immediate. We thus have to construct the bijections
$\alpha[U,V]:{\cal F}[U,V]\to F[U,V]$, which is done as follows:
Suppose that $G\in{\cal F}[U,V]$ has $k$ inner empty vertices.
Label these vertices with $1,\ldots,k$. Then define $I_l$ as the
set of legs in $4V$ that are connected with the $l$-th inner empty
vertex, together with the outer full vertices in $U$ that are
connected with the $l$-th inner full vertex. Then
$I=\{I_1,\ldots,I_k\}\in {\cal P}(U\dot\cup4V)=F[U,V]$ obviously
does not depend on the labeling of the inner empty vertices.

The construction of the inverse mapping, $\alpha^{-1}[U,V]$ is
most easily understood in the case of an example, which
generalizes in a straightforward manner to the general case: Let
us consider a partition that corresponds to a $2$-point moment
function in second order perturbation theory, i.e. $|U|=|V|=2$. To
express the moment function in the given order of perturbation
theory through truncated moment functions, see
Definition~\ref{3.1Def}, we have to sum over all partitions of $
2\times 1 + 2\times 4 = 10$ objects, as there are two exterior
vertice $U=\{u_1,u_2\} $ and two interior full vertices
$V=\{v_1,v_2\}$, each repeated four times since we consider a
partition of $U\dot\cup 4V$. Take for example the following
partition $I=\{I_1, I_2, I_3, I_4\}$ :
\begin{center}
\begin{picture}(9,3) \put(1,2){$\times$} \put(6.7,2){$\times$}
\put(3,2.2){\circle*{.08}} \put(3,2){\circle*{.08}}
\put(3.4,2.2){\circle*{.08}} \put(3.4,2){\circle*{.08}}
\put(4.6,2.2){\circle*{.08}} \put(4.6,2){\circle*{.08}}
\put(5,2.2){\circle*{.08}} \put(5,2){\circle*{.08}} \thinlines
\put(2,2.1){\oval(2.2,.5)} \put(6,2.1){\oval(2.2,.5)}
\put(4,2.225){\oval(1.5,.2)} \put(4,1.975){\oval(1.5,.2)}
\put(1.9,1.95){$I_1$} \put(3.9,2.45){$ I_2$} \put(3.9,1.5){$I_3$}
\put(5.9,1.95){$I_4$} \put(1.1,1.5){$u_1$} \put(6.8,1.5){$u_2$}
\put(3,1.5){$v_1$} \put(4.7,1.5){$v_2$} \put(3,.5){\bf Figure 2}
\end{picture}
\end{center}
Each element $I_l,\; 1\leq l\leq 4$ in the partition $I$, is now mapped to a vertex
of empty type, take e.g. $I_1$ \hspace{.2cm}
\begin{picture}(6,0)
\put(1,0){\oval(2.2,.5)}
\put(0,-.1){$\times$}
\put(1.8,.1){\circle*{.08}}
\put(1.8,-.1){\circle*{.08}}
\put(.33,-.1){$\scriptstyle u_1$}
\put(1.49,-.03){$\scriptstyle v_1$}
\put(2.2,-.1){$=$}
\put(2.65,-.1){$u_1$}
\put(3,-.06){$\times$} \put(5.05,-.05){$\bullet$}
\put(4.03,-.05){$\circ$} \put(3.15,.05){\line(1,0){.9}}
\bezier{100}(4.2,.1)(4.6,.4)(5.1,.1)
\bezier{100}(4.2,.0)(4.6,-.35)(5.1,.0)\put(5.35,-.1){$v_1$}
\end{picture}. Consequently, if there are $k$ sets in the parti-\\

\noindent tion (here $k=4$) we obtain $k$ inner empty vertices. Label them with $1,\ldots,k$. The edges of the associated graph are then given by $\cup_{l=1}^k\{ \{p,l\}, p\in I_l\}$. Thus, in the case of our example,
$\{\{u_1,1\},\{(v_1,1),1\},\{(v_1,3),1\},\{(v_1,2),2\},\{(v_1,4),3\},\{(v_2,1),2\},\{(v_2,3),3\},\{(v_2,2),4\}, \{(v_2,\linebreak4),4\},\{u_2,4\}\}$. The generalized Feynman graph is then the $\Sigma([k])$-equivalence class of the simpe graph defined by the list of edges, which removes the labeling $1,\ldots,k$.
The above partition is thus mapped to the the following
generalized Feynman graph:
\begin{center}
\begin{picture}(7,4)
\thicklines \put(0.6,2.48){$u_1$}\put(1,2.48){$\times$} \put(6.7,2.5){$\times$}\put(7,2.5)
{$u_2$}\put(3.05,2.5){$\bullet$} \put(3.05,2){$v_1$}\put(4.85,2.5){$\bullet$}
\put(4.85,2){$v_2$}\put(2.03,2.5){$\circ$} \put(5.83,2.5){$\circ$}
\put(3.94,2){$\circ$} \put(3.94,2.9){$\circ$}
\put(1.15,2.6){\line(1,0){.9}} \put(6,2.6){\line(1,0){.85}}
\put(3.4,1){$\bf Figure\;3$}
\bezier{100}(2.2,2.65)(2.6,2.95)(3.1,2.65)
\bezier{100}(2.2,2.55)(2.6,2.2)(3.1,2.55)
\bezier{100}(3.15,2.55)(3.6,2.15)(3.95,2.1)
\bezier{100}(3.15,2.65)(3.6,2.95)(3.95,3)
\bezier{100}(4.13,3)(4.5,2.95)(4.9,2.65)
\bezier{100}(4.13,2.1)(4.55,2.15)(4.9,2.55)
\bezier{100}(5,2.55)(5.4,2.15)(5.85,2.55)
\bezier{100}(5,2.65)(5.4,2.95)(5.85,2.65)
\end{picture}
\end{center}
It is easily checked that the bijections $\alpha[U,V]$ fulfill the natralness condition (\ref{3.x1eqa}). Thus we have derived the following result:

\begin{theo}
\label{3.xtheo}
The species ${\cal F}$ and $F$ are equivalent.
\end{theo}

In particular this means that we can replace a combinatorial sum in (\ref{3.9eqa}) over $F(n,m)=F[[n],[m]]$ by a combinatorial sum over the generalized Feynman graphs ${\cal F}(n,m)={\cal F}[[n],[m]]$, provided that we assign the same value to the generalized Feynman graph that has been assigned to the
corresponding (under $\alpha$) partition.

This brings us to the issue of Feynman rules. For $G\in {\cal
F}(n,m)$ clearly we have to define
 ${\cal V}[G](x_1,\ldots, x_n)=\hat {\cal
V}[\alpha^{-1}[[n],[m]](G)](x_1,\ldots,x_n)$.
 It is however
not necessary to go over the detour of partitions, as the value
can be read off the graph directly. This is done by an algorithm
called Feynman rules.
 In the given example, the Feynman rules are applied as follows:\\

\noindent \hspace{.01cm}\begin{picture}(1,0.5) \thicklines \put(0,.04){$I_1 :$}
\put(.65,.04){$u_1$}
\put(1,.04){$\times$} \put(3.05,.05){$\bullet$}
\put(2.03,.05){$\circ$} \put(1.15,.15){\line(1,0){.9}}
\bezier{100}(2.2,.2)(2.6,.5)(3.1,.2)
\bezier{100}(2.2,.1)(2.6,-.25)(3.1,.1)\put(3.25,.05){$v_1$} \put(3.6,.05)
{$\leftrightarrow$}\put(4,.05){$\langle \phi(x_1),\phi(y_1),\phi(y_1)\rangle_{\nu_0}^T$}
\put(7.5,.05){$=\langle I_1\rangle_{\nu0}^T$}
\end{picture}\\\\
\begin{picture}(1,0.5) \thicklines \put(0,.04){$I_2:$}
\put(.65,.04){$v_1$}
\put(1,.04){$\bullet$} \put(3.05,.05){$\bullet$}
\put(2.03,.05){$\circ$} \put(1.15,.15){\line(1,0){.9}}\put(2.2,.15){\line(1,0){.9}}
\put(3.25,.05){$v_2$} \put(3.6,.05)
{$\leftrightarrow$}\put(4,.05){$\langle \phi(y_1),\phi(y_2)\rangle_{\nu_0}^T$}
\put(6.5,.05){$=\langle I_2\rangle_{\nu0}^T$}
\end{picture}\\\\
\begin{picture}(1,0.5) \thicklines \put(0,.04){$I_3:$}
\put(.65,.04){$v_1$}
\put(1,.04){$\bullet$} \put(3.05,.05){$\bullet$}
\put(2.03,.05){$\circ$} \put(1.15,.15){\line(1,0){.9}}\put(2.2,.15){\line(1,0){.9}}
\put(3.25,.05){$v_2$} \put(3.6,.05)
{$\leftrightarrow$}\put(4,.05){$\langle \phi(y_1),\phi(y_2)\rangle_{\nu_0}^T$}
\put(6.5,.05){$=\langle I_3\rangle_{\nu0}^T$}
\end{picture}\\\\
\begin{picture}(1,0.5) \thicklines \put(0,.04){$I_4 :$}
\put(.65,.04){$u_2$}
\put(1,.04){$\times$} \put(3.05,.05){$\bullet$}
\put(2.03,.05){$\circ$} \put(1.15,.15){\line(1,0){.9}}
\bezier{100}(2.2,.2)(2.6,.5)(3.1,.2)
\bezier{100}(2.2,.1)(2.6,-.25)(3.1,.1)\put(3.25,.05){$v_2$} \put(3.6,.05)
{$\leftrightarrow$}\put(4,.05){$\langle \phi(x_2),\phi(y_2),\phi(y_2)\rangle_{\nu_0}^T$}
\put(7.5,.05){$=\langle I_4\rangle_{\nu0}^T.$}
\end{picture}
\vspace{.5cm}

\noindent Hence,
\begin{equation}
{\cal V}(G)=\int_\Lambda\int_\Lambda \langle\phi(x_1),\phi(y_1),\phi(y_1)\rangle^T_{\nu_0}
\left(\langle\phi(y_1),\phi(y_2)\rangle_{\nu_0}^T\right)^2\langle\phi(y_2),\phi(y_2),\phi(x_2)
\rangle_{\nu_0}^T\, dy_1dy_2
\end{equation}
This can be generalized as follows: Associate to each inner full vertex from $[m]$ an inegration variable $y_1,\ldots,y_m$. To the outer full vertices in $[n]$, the values $x_1,\ldots,x_n$ are assigned.
The connection between graphs and moments of functional
measures is established through identification \vspace{.5cm}
\begin{center}
\begin{picture}(14,1.5)
\thicklines \put(4.3,.5){\line(1,0){.8}} \put(5.08,.4){$\circ$}
\put(5.1,.55){\line(-2,1){.8}} \put(5.25,.5){\line(1,0){.7}}
\put(5,.9){$\cdots$} \put(4,.4){$w_1$} \put(6,.4){$w_q$}
\put(4,1.1){$w_2$} \put(6.45,.6){$ \leftrightarrow$}
\put(7,.6){$\langle\phi(z_1),\cdots,\phi(z_q)\rangle_{\nu_0}^T$}
\put(14.45,.6){$(24)$}
\end{picture}
\setcounter{equation}{24}
\end{center}where $w_1,\ldots, w_q\in [n]+4[m]$ and $z_1,\ldots z_q$ are the
integration/external variables associated to the full vertices to
which $w_1,\ldots,w_q$ belong. Namely , each empty inner vertex in
a generalized Feynman graph,
 with $q$ edges connected to the legs of the full vertices labeled by $w_1,...,w_q,$ is
 evaluated by the truncated moment
$\langle\phi(z_1)\cdots\phi(z_q)\rangle_{\nu_0}^T.$ Following
 this rule we obtain the general description of the evaluation of generalized Feynman
 graphs:\\
 \begin{Def}\label{3.2Def} Let $x_1,..,x_n \in \R^d$ be values assigned to the full outer vertices.
 Assign intrgration variables $y_1,...,y_m \in \R^d$  to the $m$ full inner vertices. For a generalized Feynman
 graphs $G\in {\cal F}(n,m)$,
one  proceeds as follows
:\\
\begin{enumerate}
\item For any empty inner vertex $\circ$ in $G$ with $q$ legs
(edges), we associate to this empty inner vertex $\circ,$ a
 $q$-point truncated moment function with
arguments given by the ${x_l}'s$ and ${y_j}'s$ corresponding to
the outer and inner full vertices that are connected by an edge to
that empty vertex; \item Multiply all the truncated moment
functions obtained in this fashion; \item Integrate this product
w.r.t each variable $y_j$ (interaction vertex) appearing in this
$q-$point truncated moment function over $\Lambda$.
\end{enumerate}
Denote by
${\cal V}(G)$ the obtained value.
\end{Def}
By combination of the above definition and Theorem \ref{3.xtheo}, one obtains :
\begin{equation}
\int_{{
\Lambda}^m}\int_{C^{\infty}_{temp}}
\phi(x_1)...\phi(x_n)\phi^4(y_1)...
\phi^4(y_m)d\nu_0(\phi)dy_1...dy_m=\displaystyle\sum_{G\in {\cal
F}(n,m)}{\cal V}(G).
\end{equation}
\begin{theo}\label{3.2theo}The perturbation series (\ref{2.8eqa}) can be expressed as follows:
\begin{equation}
S_{n,\Lambda}^\lambda(x_1,\ldots,x_n)=\displaystyle\sum_{m=0}^{\infty}\frac{(-\lambda)^m}{m!}
\sum_{G\in{\cal F}(n,m)}{\cal V}(G).
\end{equation}
\end{theo}
\noindent  In particular this procedure applies to the case when
$\nu_0$ is Gaussian. One recovers the conventional Feynman graphs:
\begin{ex}
{\rm Let $\nu_0$ be a Gaussian measure. Then the only
non$-$vanishing truncated correlation function is the two point
function, or in graphical expression, the only empty inner vertex
which does not lead to a zero evaluation of the generalized
Feynman graph is
\hspace{.45cm}\begin{picture}(0,0)\put(0,.1){\line(1,0){.3}}\put(.3,.0){$\circ
$} \put(.42,.1){\line(1,0){.3}}\put(.75,.05){$.$}\end{picture}\\A
non$-$vanishing graph, say of $\phi^2-$theory for a two point
function, is thus of the kind : }
\end{ex}
\begin{center}
\begin{picture}(12,1.5)

\thicklines \put(.05,.385){$\times$}
\put(.21,.49){\line(1,0){.25}} \put(.44,.385){$\circ$}
\put(.61,.49){\line(1,0){.25}} \put(.84,.385){$\bullet$}
\put(1.01,.49){\line(1,0){.25}} \put(1.24,.385){$\circ$}
\put(1.41,.49){\line(1,0){.25}} \put(1.64,.385){$\bullet$}
\put(1.82,.49){\line(1,0){.25}} \put(2.05,.385){$\circ$}
\put(2.22,.49){\line(1,0){.28}} \put(2.41,.38){$\times$}
\put(1.24,.085){$\circ$} \put(1.24,.685){$\circ$}
\bezier{100}(.9,.55)(1.05,.75)(1.25,.8)
\bezier{100}(.9,.45)(1.05,.25)(1.25,.2)
\bezier{100}(1.42,.8)(1.62,.75)(1.75,.55)
\bezier{100}(1.42,.2)(1.62,.25)(1.75,.45) \put(3,.385){$=$}
\put(3.79,.385){$\times$} \put(4.6,.385){$\bullet$}
\put(5.4,.385){$\bullet$} \put(6.04,.385){$\times$}
\bezier{6}(3.95,.49)(4.3,.49)(4.65,.49)
\bezier{6}(4.7,.49)(5.05,.49)(5.4,.49)
\bezier{6}(5.5,.49)(5.85,.49)(6.2,.49)
\bezier{7}(4.7,.55)(5.125,1)(5.45,.53)
\bezier{7}(4.7,.45)(5.125,0)(5.45,.45) \put(6.6,.45){with}
\put(7.7,.45){$.....$}\put(8.5,.37){$=$}
\put(9.01,.49){\line(1,0){.52}} \put(9.56,.4){$\circ$}
\put(9.7,.49){\line(1,0){.5}} \put(5,-1){\bf Figure 4}
\end{picture}
\end{center}
\vspace{1cm} where the second graph in Figure 4 is the
conventional Feynman graph which is being evaluated by putting a
two point function for each line (conventional Feynman rules). In
this situation, the two$-$point function is also called Euclidean
propagator.
\section{Wick ordering by avoiding self-contractions}
As an example, let us first have a look at one interaction vertex
$(inner\; full\; vertex)$ in $\phi^4-$interaction case. In
Definition \ref{3.2Def} the following situations are not excluded
:\begin{center}\begin{picture}(7,2)\thicklines \put(-.435
,.475){$\circ$} \put(.7 ,.45){$\circ$} \put(.64
,-.65){$\circ$}\thicklines
\put(2.7,-.1){$\bullet$}\put(2.65,0){\line(-1,0){.5}}\put(2
,-.1){$\circ$} \bezier{100}(2.8,-.02)(2,.35)(2.625,.7)
\bezier{100}(2.75,-.02)(3.5,.35)(2.79,.741)
\put(2.65,.6){$\circ$}\put(2.65,0){\line(1,0){.6}}
\put(3.25,-.1){$\circ$} \thicklines
\put(.15,0){\line(-1,1){.5}}\put(.13,-.08){\line(-1,-1){.6}}
\put(-.6,-.81){$\circ$}\put(.15,-.08){\line(1,1){.6}}\put(.7,.45){$\circ$}
\put(.13,-.14){$\bullet$}\put(.15,0){\line(1,-1){.5}}\thicklines
\put(5.15,-.1){$\bullet$}\put(5.15,0){\line(-1,0){.5}}\put(4.5
,-.1){$\circ$}\put(5.22,.01){\line(0,1){.7}}\put(5.15
,.65){$\circ$} \bezier{100}(5.18,0)(4.6,.35)(5.15,.75)
\bezier{100}(5.3,.03)(5.8,.35)(5.31,.745)
\thicklines\put(6.15,-.1){$\bullet$}\put(6.18,0){\line(0,1){.68}}\put(6.15
,.6){$\circ$}\put(6.28,0){\line(0,1){.67}}
\bezier{100}(6.15,0)(5.6,.4)(6.18,.75)
\bezier{100}(6.23,-.015)(6.7,.4)(6.3,.742) \put(2.4,-1.5){$\bf
Figure\;5$}
\end{picture}\end{center}\vspace{1.5cm}In each of these parts of a generalized Feynman
graph we have that one empty vertex is only connected to one and
the same interaction vertex. In this situation we say, that a
self-contraction occurs at the interaction vertex. Generally we
say, that a self-contraction occurs in a generalized Feynman
graph, if there exists in this graph, an empty inner vertex
connected with one and only one interaction vertex (full inner
vertex). \noindent Generally speaking, such self-contractions are
not problematic for measures $\nu_0$ with $C^{\infty}(\R^d)-$paths
where the truncated Schwinger functions are continuous functions.
In the case however where ${\langle \phi(x_1),...,\phi(x_n)
\rangle}^T_{\nu_0}$ is a function with singularities if
$x_j=x_l,\; j\neq\; l,$ a self$-$contraction leads to a
uv-divergence, cf. e.g.\vspace{.3cm}
\begin{center}
\begin{picture}(8,.15)
\thicklines
\put(2.15,-.1){$\bullet$}\put(2.15,0){\line(-1,0){.5}}\put(1.5
,-.1){$\circ$}\put(1.5
,0){\line(-1,0){.3}}\bezier{100}(2.18,0)(1.8,.25)(2.17,.7)\put(2.15
,.6){$\circ$}\bezier{100}(2.3,0)(2.6,.25)(2.3,.7)\put(2.21,0){\line(0,1){.65}}
\put(1,.02){$\times$} \put(1,-.2){$\times$} \put(2.9,-.1){$=$}
\thicklines \put(5.15,-.1){$\circ$} \put(5.15,0){\line(-1,0){.5}}
\put(4.45,.02){$\times$} \put(4.45,-.2){$\times$}
\put(5.25,0){\line(1,0){.3}} \put(5.5,-.1){$\bullet$}
\put(5.9,0){$.$}\put(6.2,0){${\langle\phi^3(y)\rangle}^T_{\nu_0}$}
\put(8,0){$.$}\put(11.5,0){$(27)$}
\end{picture}
\setcounter{equation}{27}
\end{center}
 It is thus desirable (even though this often
does not resolve the problem of UV divergences completely) to
avoid self-contractions. The procedure which is used to do so is
called Wick-ordering.\\It is known that Wick-ordering can be done
in the Gaussian case avoiding self-contractions of the form
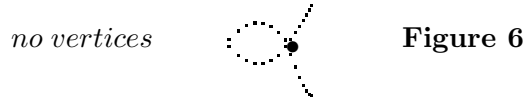
\begin{center}
\begin{picture}(12,1.5)
\thicklines \bezier{7}(4.7,.55)(5.125,1)(5.45,.53)
\bezier{7}(4.7,.45)(5.125,0)(5.45,.45)
\bezier{7}(5.45,.45)(5.7,.8)(5.8,1)
\bezier{7}(5.45,.55)(5.7,-.2)(5.8,-.2) \put(5.45,.35){$\bullet$}
\put(1.7,.45){$\;no\;vertices\;$} \put(7,.45){\bf Figure 6}
\end{picture}
\end{center} Here
we develop a formalism for Wick$-$ordering in the general case.
Because of the one-to-one correspondance between the empty inner
vertices of a generalized Feyman graph, and the elements of the
related partition, note that a self-contraction in a generalized
Feynman graph occurs, if in the related partition $I$ there is a
subset $I_l$ that is contained in one of the $m$ sets standing for
the interaction vertices and containing four points each. We can
thus formulate the problem on the level of partitions and truncated
moment functions.  \\Let $Y=Y(\phi)$ be a '' sufficiently integrable
'' random variable$-Y \in L^p(\nu_{0})$ for some $p \geq 1$ would
do. As in Equation (\ref{3.1eqa}) one has
:\begin{equation}\label{4.1eqa}{\langle
\phi(u_1)...\phi(u_n)Y\rangle}_{\nu_0}=\sum_{I \in {\cal
P}(\{1,...,n+1\})} \prod_{I_l\in I}{\langle
I_l\rangle}^T_{\nu_0}\end{equation} where we use the convention that
on the right hand side we replace $\phi(u_{n+1})$ by
$Y(\phi).$\\(\ref{4.1eqa}) is to underline the analogy with
(\ref{3.1eqa}). One can also rewrite (\ref{4.1eqa}) in the more
explicit form :
\begin{eqnarray}\label{4.2eqa}{\langle
\phi(u_1)...\phi(u_n)Y\rangle}_{\nu_0}&=&\sum_{I=\{I_1,...,I_k\}
\in\; {\cal P}(\{1,...,n\})} \sum_{j=1}^k {\langle I_j,
Y\rangle}^T_{\nu_0} \prod^k_{l=1,\; l\neq j}{\langle I_l
\rangle}^T_{\nu_0}\; \nonumber \\&+& \langle Y\rangle_{\nu_0}
\sum_{I=\{I_1,...,I_k\} \in\; P(\{1,...,n\})}\prod^k_{l=1}{\langle
I_l \rangle}^T_{\nu_0}\end {eqnarray}where $\langle
I_j,Y\rangle^T_{\nu_0}=\langle
\phi(u_{i_1}),...,\phi(u_{i_p}),Y\rangle^T_{\nu_0}$ if
$I_j=\{i_1,...,i_p\},$ and $\; \langle
I_l\rangle^T_{\nu_0}=\langle
\phi(u_{j_1}),...,\phi(u_{j_r})\rangle^T_{\nu_0}\;$ for $\;
I_l=\{j_1,...,j_r \}.$ It is clear, that there are lots of
self$-$contractions of the type $\langle I_l \rangle^T_{\nu_0}$ on
the right hand side of (\ref{4.2eqa}). In fact, there is only one
single term free of self$-$contractions: it is $\langle
\phi(u_1),...,\phi(u_n),Y\rangle^T_{\nu_0}.$ If Wick$-$ordering is
to remove all self$-$contractions, one thus has to set
:\begin{equation}\label{4.3eqa}\langle:\phi(u_1)...\phi(u_n):Y\rangle_{\nu_0}=\langle\phi(u_1),
...,\phi(u_n),Y\rangle^T_{\nu_0}\end{equation} Using
(\ref{4.2eqa}) and (\ref{4.3eqa}) recursively, one obtains :
\begin{Def}\label{4.2Def}
The Wick$-$ordering monomial is recursively defined by the
equation
\begin{equation}\label{4.4eqa}\begin{array}{lll}
:\phi(u_1)...\phi(u_n):=
    \phi(u_1)...\phi(u_n)&-{\displaystyle \sum_{\{I_1,...,I_k\}\in\;{\cal P}(\{1,...,n\})\;
    k> 1}}
    \displaystyle\sum_{j=1}^k\;
                        :I_j:\prod_{l=1,\; l\neq j}^k {\langle I_l\rangle}^T_{\nu_0}&\\\\
                         &-\displaystyle\sum_{\{I_1,...,I_k\}\in\; {\cal P}(\{1,...,n\})}
                         \prod_{l=1}^k
                         {\langle
I_l\rangle}^T_{\nu_0} &
\end{array}
\end{equation}
where $:I_j:$ stands for $:\phi(u_{i_1})...\phi(u_{i_p}):$ if
$I_j=\{i_1,...,i_p\},$  and $\; \langle
I_l\rangle^T_{\nu_0}=\langle
\phi(u_{j_1}),...,\phi(u_{j_r})\rangle^T_{\nu_0}\;$ for $\;
I_l=\{j_1,...,j_r \}.$
\end{Def}
\noindent We have to show that this definition in fact solves the
problem of self-contractions.
\begin{theo}\label{4.1theo}
If in Theorem~\ref{3.2theo} one replaces the interaction $\phi^4$
with the interaction $:\phi^4:,$ one obtains the same Feynman
graphs and rules with the only difference that all Feynman graphs
which have a self$-$contraction at an interaction vertex are
omitted.
\end{theo}
 To prove the above theorem, it is sufficient to prove the
following more general case :$${\langle
:\phi(y_1)...\phi(y_{p_1})::\phi(y_{p_1+1})...\phi(y_{p_1+p_2}):...:\phi(y_{p_1+...+p_
{m-1}+1})...\phi(y_{p_1+...+p_m}):
\phi(x_1)...\phi(x_n)\rangle}_{\nu_0}$$
\begin{equation}\label{4.5eqa}=\displaystyle\sum_{\begin{array}{ll}I\; \in\;
{\cal P}_{sc}(J_1,...,J_m;Y)\\
I=\{I_1,...,I_k\}\end{array}}\prod_{j=1}^k{\langle
I_j\rangle}^T_{\nu_0}\end{equation}where $J_1,J_2,...,J_m$ are
respectively the subsets $\{1,...,{p_1}\}$,
$\{{p_1+1},...,{p_1+p_2}\},$ $..., \{p_1+...+\linebreak
p_{m-1}+1,..., {p_1+...+p_m}\},$ $Y=\{1,...,n\}$ and ${\cal
P}_{sc}(J_1,...,J_m;Y)$ is the set of all partitions
$I=\{I_1,...,I_k\}$ of the elements of $J_1\cup...\cup J_m,Y$ such
that $I_l\nsubseteq J_q$ for all $1\leq l\leq k$ and $1\leq q\leq
m.$
\\\\ \noindent{\bf Proof of (\ref{4.5eqa}):}\\Let
$q=\sum_{i=1}^m|J_i|$ where $|J_i|$ is the cardinal number of the
set $J_i ,$ for example $|J_1|=p_1.$ The objective is to prove
(\ref{4.5eqa}) by induction on q. If $q=0 ,$ $J_i=\emptyset\;
\forall\; i ,$ then the statement holds simply by definition of
truncation.\\ Let now $q> 0$ and we assume that (\ref{4.5eqa})
holds up to $q-1$. The induction step can be seen as follows: We
first apply the definition of Wick ordering to $J_m$ where without
loss of generality one may assume $J_m\not=\emptyset$,
\\\\$\langle:J_1:...:J_{m-1}:\;:J_m:Y\rangle_{\nu_0}=
\langle:J_1:...:J_{m-1}:J_m Y \rangle_{\nu_0}
-\langle:J_1:...:J_{m-1}:
\\\\ \times\left(\displaystyle\sum_{I=\{I_1,...,I_k\}\; \in\; {\cal P}(J_m),\; k\;>\;
1}\;\sum_{j=1}^k\; :I_j:\; \prod_{\begin{array}{ll}l=1\\l\neq
j\end{array}}^k {\langle I_l
\rangle}^T_{\nu_0}+\sum_{I=\{I_1,...,I_k\}\; \in\;{\cal P}(J_m)}\;
\prod_{j=1}^k{\langle I_j\rangle}^T_{\nu_0}\right)Y
\rangle_{\nu_0}$\\\\$\begin{array}{ll}=\langle:J_1:...:J_{m-1}:J_m
Y \rangle_{\nu_0}-\displaystyle\sum_{\begin{array}{ll}I\; \in\;
{\cal P}(J_m); k\;>\; 1\\ I=\{I_1,...,I_k\}\end{array}
}\sum_{j=1}^k\;\langle:J_1:...:J_{m-1}:\; :I_j: Y \rangle_{\nu_0}
\\ \qquad\qquad\qquad\qquad\qquad\qquad\qquad\qquad\qquad\times
\displaystyle\prod_{l=1,\; l\; \neq\; j}^k \; {\langle I_l
\rangle}^T_{\nu_0}-\langle
J_m\rangle_{\nu_0}\;\langle:J_1:...:J_{m-1}:\;
Y\rangle_{\nu_0}~(33)\end{array}$\\ \setcounter{equation}{33}Using the hypothesis of induction
gives :\begin{equation}\label{4.6eqa}{\langle
:J_1:....:J_{m-1}:J_mY\rangle}_{\nu_0}=\displaystyle\sum_{\begin{array}{ll}I\;
\in\; {\cal P}_{sc}(J_1,...,J_{m-1};J_m,Y)\\ I=\{I_1,...,I_k\}
\end{array}}\prod_{j=1}^k\langle
I_j\rangle^T_{\nu_0}\end{equation}because
$\displaystyle\sum_{i=1}^{m-1} |J_i| < q.$
 On the right hand side of (\ref{4.6eqa}) there is the
sum over all the partitions of the elements of $J_1,...,J_m, Y$
which do not have self$-$contractions in $J_1,...,J_{m-1}$.
On the other hand we
have\begin{equation}\label{4.7eqa}\begin{array}{ll}&
\displaystyle\sum_{I=\{I_1,...,I_k\}\; \in\;{\cal P}(J_m),\;
k\;>\; 1}\displaystyle\sum_{j=1}^k\;\langle:J_1:...:J_{m-1}:\;
:I_j:Y \rangle_{\nu_0}\; \prod_{l=1,\; l\; \neq\; j}^k \; {\langle
I_l \rangle}^T_{\nu_0}\\\\&=\displaystyle\sum_{I=\{I_1,...,I_k\}\;
\in\; {\cal P}(J_m),\; k\;>\; 1}\;\sum_{j=1}^k
\sum_{\begin{array}{ll}\tilde I\;\in {\cal
P}_{sc}(J_1,...,J_{m-1},I_j;Y)\\ \tilde
I=\{\tilde{I_1},...,\tilde{I_p}\}
\end{array}}\displaystyle\prod_{r=1}^p{\langle
\tilde{I_r}\rangle}^T_{\nu_0}\; \prod_{l=1,\; l\neq\; j}{\langle
I_l\rangle}^T_{\nu_0}
\end{array}\end{equation}
because $\;\displaystyle\sum_{i=1}^{m-1}|J_i|+|I_j| <\; q\;
\forall\; j=1,...,k.$ Using the hypothesis of induction, one
obtains
\begin{equation}\label{4.8eqa}\begin{array}{ll}& \langle
J_m\rangle_{\nu_0}{\langle:J_1:...:J_{m-1}:Y\rangle}_{\nu_0}\\\\ &
=\displaystyle\sum_{I=\{I_1,...,I_k\}\; \in \;{\cal P}(J_m)}\;
\prod_{l=1}^k\; \langle
I_l\rangle^T_{\nu_0}\sum_{\begin{array}{ll}\tilde{I}\in\; {\cal
P}_{sc}(J_1,...,J_{m-1};Y)\\
\tilde{I}=\{\tilde{I_1},...,\tilde{I_p}\}\end{array}}
\displaystyle\prod_{r=1}^p \langle \tilde{I_r}\rangle^T_{\nu_0}
\end{array} \end{equation}
In
(\ref{4.7eqa}), the sum is over all the partitions of the elements
of $J_1,...,J_m, Y$ which do not have self$-$contractions in
$J_1,...,J_{m-1}$ but have self$-$contractions in $J_m$ and such
that the union of the elements of the every partition which are
included in $J_m,$ is strictly included in $J_m.$\\(\ref{4.8eqa})
is the sum over all the partitions of the elements of
$J_1,...,J_m, Y$ which do not have self$-$contractions in
$J_1,...,J_{m-1}$ but have self$-$contractions in $J_m$ and such
that the union of the blocks of the every partition which are
included in $J_m$ is exactly $J_m.$ Hence :\\ $\langle
:J_1:...:J_m:Y\rangle_{\nu_0}=(\ref{4.6eqa})-(\ref{4.7eqa})-(\ref{4.8eqa})=$
the sum over all the partitions of $(J_1,...,J_m,Y)$ which do not
have self$-$contractions in $J_1,...,J_m$.\prend \\ The following
clarifies the relation between our graphical notion of Wick
ordering and orthogonal decompositions of Wiener-It\^o-Segal type:
\begin{theo}\label{4.2theo}
The Wick ordered monomials are orthogonal with respect to
$L^2(\nu)$ if and only if the measure $\nu$ is Gaussian. In
particular, Wick ordering in the sense of Definition \ref{4.2Def}
leads to a chaos (orthogonal) decomposition of $L^2(\nu)$ if and
only if $\nu$ is a Gaussian measure.
\end{theo}
\noindent{\bf Proof.} Coincidence of Wick ordering and the
orthogonal decomposition is well-known for Gaussian measures. If
 $\nu$ is not a Gaussian measure, then there exist $ u_1,...,u_r \in \R^d,\; r\geq
 3$  such that $\langle\phi(u_1),...,\phi(u_r)\rangle^T \neq 0.$ Let $r$
 be the minimal number with this property.
 Then $$\langle:\phi(u_1)::
 \phi(u_2)...\phi(u_r):\rangle_{\nu}=\langle\phi(u_1),
 \phi(u_2),...,\phi(u_r)\rangle_{\nu}^T.$$Hence $\langle:\phi(u_1)::
 \phi(u_2)...\phi(u_r):\rangle_{\nu}\neq 0.$ Hence the Wick
 ordered monomials are not orthogonal with respect to
 $L^2(\nu).$\prend

\section{Expansion of the free energy into connected Feynman graphs}

The topic of this section is a classical problem of statistical
mechanics, namely the expansion of the free energy $\ln
\Xi_\Lambda(\lambda)$ of a statistical mechanics system into
connected Feynman graphs. Here we give a new and measure
independent proof of a very general linked cluster theorem:
\begin{theo}\label{5.1theo}
Only connected generalized Feynman graphs contribute to the
perturbation expansion of the free energy, i.e., in the sense of
formal power series in $\lambda$
\begin{equation}\label{5.1eqa}
\ln \Xi_\Lambda(\lambda)=\sum_{m=1}^\infty{(-\lambda)^m\over
m!}\sum_{G \in{\cal F}(0,m)\atop G~{\rm connected}}{\cal
V}_\Lambda(G)
\end{equation}
where ${\cal V}_\Lambda(G)$ is the value associated with $G$
according to the Feynman rules given in Definition~\ref{3.2Def}
\end{theo} \noindent The remainder
of this section is dedicated to the proof of this theorem. We would like to remark that this theorem also
holds
when the interaction is Wick ordered, as the proof we give here
carries over word by word.

Let us first observe that $\Xi_\Lambda(\lambda)$ is the Laplace
transform of the random variable $V_\Lambda$, hence it is the
generating functional for the moments of $V_\Lambda$. It follows
>from the basic linked cluster theorem~\ref{3.1theo} (that is valid
also for the Laplace transform) that $\ln \Xi_\Lambda(\lambda)$ is
the generating functional of the truncated moments of $V_\Lambda$,
i.e.
\begin{equation}\label{5.2eqa}
\ln\Xi_\Lambda(\lambda)=\sum_{m=1}^\infty {(-\lambda)^m\over
m!}\langle \underbrace{V_\Lambda,\ldots,V_\Lambda}_{m\rm~times} \rangle^{T}_{\nu_0}.
\end{equation}
Furthermore, using induction and Fubini's theorem, it
is easy to show that
\begin{equation}\label{5.3eqa}
\langle \underbrace{V_\Lambda,\ldots,V_\Lambda}_{m\rm~times} \rangle^{T}_{\nu_0}=\int_{\Lambda^m}\langle
\phi^4(y_1),\cdots,\phi^4(y_m)\rangle^{T}_{\nu_0}dy_1\cdots dy_m
\end{equation}
 \noindent The following theorem is the main technical step. It expands the
truncated moments $\langle J_1,... ,J_n\rangle^{(T)}_{\nu_0}$ into  truncated moments $\langle
\phi(u_1),...,\phi(u_m)\rangle^{T}_{\nu_0} :$
\begin{theo}\label{5.2theo} The following identity holds:
\begin{equation}\label{5.5eqa} \langle
J_1,...,J_n \rangle^{T}_{\nu_0}=\sum_{I=\{\; I_1,...,I_k\}\in\;
P_c(\bigcup_{l=1}^n J_l) }\prod_{l=1}^k \langle
I_l\rangle_{\nu_0}^T
\end{equation} where $P_c(\bigcup_{l=1}^n J_l)$
is the all elements $I=\{\;I_1,...,I_k\}\in\; P(\bigcup_{l=1}^n
J_l)$ such that $\forall\; 1\leq J_{j_1}<\;...<\;J_{j_l}\leq n$
with $\;l<~n,$ and $\forall \; 1\leq i_1<...<i_r\leq k,$ one has
$\bigcup_{q=1}^l J_{j_q}\neq\bigcup_{p=1}^r I_{i_p}.$
\end{theo} \noindent{\bf Proof.} To prove (\ref{5.5eqa}) it is
sufficient to prove that the right hand side of (\ref{5.5eqa})
fulfills the defining equation (\ref{3.1eqa}), i.e.,
\begin{equation}\label{5.6eqa} \langle
J_1...J_n\rangle_{\nu_0}=\sum_{I=\{I_1,...,I_k\}\in\;P(\{1,...,n\})}
\prod_{l=1}^k \sum_{\begin{array}{ll} Q^l\in\;P_c(\bigcup_{q\in\;
I_l}
J_q)\\Q^l=\{Q^l_1,...,Q^l_{r_l}\}\end{array}}\prod_{a=1}^{r_l}\langle
Q^l_a\rangle^T_{\nu_0} \end{equation}for any collection
$J_1,...,J_n$. Expanding the left hand side into truncated
functions, we get
$$\sum_{\begin{array}{ll}\alpha\in\; P(\bigcup_{l=1}^n J_l)\\
\alpha=\{\alpha_1,...,\alpha_k\}\end{array}}\prod_{j=1}^k \langle
\alpha_j\rangle^T_{\nu_0}=\sum_ {\begin{array}{ll} I\in\;
P(\{1,...,n\})\\I=\{I_1,...,I_k\}\end{array}}\sum_{\begin{array}{ll}Q^1\in\;
P_c(\bigcup_{q\in\;
I_1}J_q)\\Q^1=\{Q^1_1,...,Q^1_{r_1}\}\end{array}}\prod_{a=1}^{r_1}
\langle Q^1_a\rangle^T_{\nu_0}...$$ $$...
\sum_{\begin{array}{ll}Q^k\in\; P_c(\bigcup_{q\in\;
I_k}J_q)\\Q^k=\{Q^k_1,...,Q^k_{r_k}\}\end{array}}\prod_{a=1}^{r_k}
\langle Q^k_a\rangle^T_{\nu_0}$$ $$=\sum_ {\begin{array}{ll}
I\in\;
P(\{1,...,n\})\\I=\{I_1,...,I_k\}\end{array}}\sum_{\begin{array}{ll}Q^1\in\;
P_c(\bigcup_{q\in\;
I_1}J_q)\\Q^1=\{Q^1_1,...,Q^1_{r_1}\}\end{array}}...\sum_{\begin{array}{ll}Q^k\in\;
P_c(\bigcup_{q\in\;
I_k}J_q)\\Q^k=\{Q^k_1,...,Q^k_{r_k}\}\end{array}}$$
\begin{equation}\label{5.7eqa}\prod_{a=1}^{r_1}\langle
Q^1_a\rangle^T_{\nu_0}\quad...\quad\prod_{a=1}^{r_k} \langle
Q^k_a\rangle^T_{\nu_0}\end{equation} Let us focus on the case
$\sharp J_l=4$, for simplicity (these considerations however carry
over to the general case). Then there is a one to one
correspondence between generalized Feynman graphs with
$\phi^4$-interaction vertices and partitions of $J_1,\ldots,J_m$,
cf. section 3. Following the prescription of the equivalence
between partitions and Feynman graphs, it is not difficult to see
that the associated generalized Feynman graphs to
$P_c(\bigcup_{l=1}^n J_l)$ are just the connected Feynman
generalized graphs. Hence, the first sum on the right hand side of
(\ref{5.7eqa}) is a sum over all possible connected components of
the graph with "vertices" $J_1,\ldots,J_m$. The remaining sums
then give all possibilities of graphs which, have exactly the
given connected components. Hence on both sides of (\ref{5.7eqa})
we get a sum over all generalized Feynman graphs or, equivalently,
over all partitions.
 \prend

As explained in the proof above, the interpretation of the
condition in the sum of (\ref{5.5eqa}) just means that the
associated Feynman graph is connected. Combining thus the
equations (\ref{5.2eqa})--(\ref{5.5eqa}) one obtains Theorem
\ref{5.1theo}.

\

\noindent {\bf Acknowledgements.} H.G. has been supported by the
D.F.G. project "Stochastic methods in quantum field theory". S. H.
Djah and H. Ouerdiane thank Prof. Albeverio for kind invitations
to the Institute of Applied Mathematics in Bonn. We thank the referees for putting forward a number of suggestions that considerably improved the presentation
 of this paper.

\end{document}